# Imaging Electrical Conduction through InAs Nanowires

*Ania C. Bleszynski[1], Floris A. Zwanenburg[2], Robert M. Westervelt\*[1], Aarnoud L. Roest[3],*

*Erik P. A. M. Bakkers[3], Leo P. Kouwenhoven[2]*

[1]Department of Physics and Division of Engineering and Applied Sciences,

Harvard University, Cambridge, Massachusetts 02138, USA.

[2]Kavli Institute of Nanoscience, Delft University of Technology, Delft, The Netherlands.

[3]Philips Research Laboratories, Eindhoven, The Netherlands.

\*Corresponding Author: westervelt@deas.harvard.edu

**Abstract:** We show how a scanning probe microscope (SPM) can be used to image electron flow through InAs nanowires, elucidating the physics of nanowire devices on a local scale. A charged SPM tip is used as a movable gate. Images of nanowire conductance *vs*. tip position spatially map the conductance of InAs nanowires at liquid He temperatures. Plots of conductance *vs*. back gate voltage without the tip present show complex patterns of Coulomb-blockade peaks. Images of nanowire conductance identify multiple quantum dots located along the nanowire - each dot is surrounded by a series of concentric rings corresponding to Coulomb blockade peaks. An image locates the dots and provides information about their size. The rings around individual dots interfere with each other like Coulomb blockade peaks of multiple quantum dots in series. In this way, the SPM tip can probe complex multi-dot systems by tuning the charge state of individual dots. The nanowires were grown from metal catalyst particles and have diameters ~ 80 nm and lengths 2 to 3 μm.

An explosion in research activity on semiconducting nanowires has occurred in the past decade [1-3]. The ability to control the dimensions and composition of nanowire devices shows great promise for future nano-electronics, nano-photonics, and quantum information processing. Quantum effects are naturally important due to the small size, opening new possibilities for quantum devices.

InAs nanowires are a particularly attractive system for several reasons. InAs has a large g-factor, making it useful for spintronics and quantum information processing applications. Its large bulk exciton Bohr radius $a_B$ = 34nm is comparable to the radius of the nanowires studied in this paper, producing quantum confinement. Whereas some semiconductors are known to have a surface depletion layer, the surface of InAs is known to have a charge accumulation layer. This potentially allows for very small radius nanowires that are not depleted of electrons, as well as Schottky-barrier-free contact to metallic leads.

Many recent achievements have been made in the field of semiconducting nanowires including single electron control [4-6], high performance field-effect transistors [7], and proximity induced superconductivity [8]. Progress in these fields requires an understanding of where the electrons are along the nanowire and how they flow through it. Standard transport measurements have yielded much information about the electrical properties of the wires, but averaged over the whole length of the wire [4-6].

SPM imaging allows one to locally probe the motion of electrons along the wire and modify the potential profile to locally allow or block electron transport with high spatial resolution. A variety of scanned probe techniques based on an atomic force microscope (AFM) have proven to be powerful tools capable of locally probing and manipulating low-dimensional nanoscale systems such as carbon nanotubes, GaAs quantum dots, and two-dimensional electron gases [9-15]. Nanowire imaging techniques are just being developed [16-17] and this work represents the first low-temperature scanning gate imaging measurement on semiconducting nanowires.

In this letter we present images of conductance through InAs nanowires at liquid He temperatures. Simple plots of conductance $G$ vs. backgate voltage $V_{bg}$ without the tip present show complex patterns of Coulomb blockade peaks with uneven spacings and heights. The spatial pattern of electron flow can be imaged and understood by using the SPM tip as a movable gate: the tip voltage $V_{tip}$ can locally change the electron density below. An image of electron flow through a nanowire is obtained by displaying the conductance *vs.* tip position as the tip is scanned in a plane above. Using InAs nanowire conductance images, we find that a number of quantum dots form along the nanowires. Each dot is at the center of a set of concentric rings of high conductance that correspond to Coulomb blockade conductance peaks that occur as electrons are added to that dot [11]. The rings from nearby dots overlap. By using the tip as a movable gate we can tune the charge state of each dot individually. The spacing and intensity of the rings provide information about dot size and tunneling rate, and the interference between overlapping rings gives information about the interdot coupling. These results show how a cooled SPM can be a powerful diagnostic tool for nanowire devices.

The InAs nanowires were grown in a catalytic process from small gold seed particles using metal-organic vapor phase epitaxy (MOVPE) [18]. The nanowires have diameters ~80 nm and lengths ~2-3 μm. After growth, the InAs nanowires are transferred onto a conducting p+ silicon substrate capped with a 250nm thick $SiO_2$ insulating layer. The silicon substrate acts as a back gate that can tune the number of charge carriers in the wire through an applied voltage. Electron beam lithography is used to define electrodes ~ 2 μm apart and 110 nm of Ti/Al is subsequently deposited to form the contacts. Figure 1A shows an SEM picture of a contacted InAs wire.

We use a home-built liquid-helium cooled scanning probe microscope (SPM) to image electrical conduction through the nanowires. As schematically shown in Fig. 1B, we laterally scan a conducting SPM tip along a plane above the nanowire and record the nanowire conductance *vs.* tip position to form an image [11]. The conducting tip allows us to gate the nanowire *locally*, whereas the backgate gates the nanowire *globally*. Images of conduction through the nanowire were obtained by displaying the nanowire conductance $G$ as the tip was scanned above with a fixed tip and backgate voltages. The tip voltage $V_{tip}$ creates a dip or peak in the electron density below. For an open nanowire, one could image electron flow by using tip-induced density change to scatter electrons, thereby changing $G$. However, for a quantum dot in the Coulomb blockade regime a different pattern is observed. An image of the dot shows a series of concentric rings of high conductance that correspond to the conductance peaks that occur as electrons are added to the dot. This Coulomb blockade imaging technique has been used to image a one-electron GaAs quantum dot [11] and multi-electron quantum dots formed in carbon nanotubes [10]. We apply this imaging technique to semiconducting nanowires for the first time.

Figures 2A-B show plots of nanowire conductance $G$ *vs.* back gate voltage $V_{bg}$ for two InAs nanowire devices, D1 and D2; each plot shows an irregular series of peaks.

The corresponding SPM conductance images in Figs. 2C-D allow us to understand how these plots came about by identifying multiple quantum dots along each nanowire and by imaging their location and size. For the plots in Figs 2A-B, the wires are near pinch-off and they exhibit irregular Coulomb-blockade oscillations indicative of multiple dots in series [19-21]. For a single dot, one would expect to see isolated Coulomb peaks with a regular spacing. For multiple dots in series with inter-dot coupling and varying capacitances to the gate, "stochastic" Coulomb blockade peaks are expected [21]. Without additional information, it is impossible to say what along the wire is giving rise to these irregular Coulomb oscillations. Therefore we have used imaging to spatially probe the nanowire on a local scale in order to elucidate the local electrostatic fluctuations that give rise to quantum dot behavior.

SPM conductance images of devices D1 and D2 are shown in Figs. 2C and 2D, respectively. Nested rings of peaked conductance occur about three positions along the nanowire in Fig. 2C and about two positions in Fig. 2D. As described in [10-11], each ring corresponds to a Coulomb conductance peak of the quantum dot at the ring's center. The charge induced by the tip on a single dot is given by

$$q_{ind}(r_{t-d}, V_{t-d}) = C_{t-d}(r_{t-d}) * V_{t-d} \quad (1)$$

where $r_{t-d}$ is the distance between the tip and the dot, $C_{t-d}$ is the capacitance between the tip and the dot (assuming a conducting dot with a fixed geometry), and $V_{t-d}$ is the voltage difference between tip and dot, including effects of the contact potential and the dot's capacitance to ground. Because $C_{t-d}$ changes with tip position, the induced charge $q_{ind}$ can be controlled by either the tip voltage $V_{tip}$ or the tip position $r_{t-d}$. If one were to plot

*G* vs. $r_{t-d}$, a conductance peak would occur every time the charge in the dot changes by one electron. In images, the conductance peaks take the form of closed rings centered on the dot that are contours of constant tip-to-dot capacitive coupling $C_{t-d}$. When the tip is between two rings, the dot charge remains constant at an integral multiple of *e*.

The conductance images for devices D1 and D2 in Figs. 2C-D show that the complex conductance plots in Figs. 2A-B were caused by multiple quantum dots in series. In Fig. 2C, three sets of concentric rings indicate the presence of three quantum dots at locations indicated by the black dots superimposed on the image. The rings surrounding the middle dot in D1 are more closely spaced than those surrounding the other two dots, indicating that the center dot is larger than the other two. In Fig. 2D, two sets of concentric rings indicating the presence of two quantum dots, whose locations are again marked by black dots. In both Figs. 2C and 2D, the rings are elongated along an axis perpendicular to the wire due to a slight screening of the tip by the metal contacts. Formation of the quantum dots is presumably due to local potential fluctuations or defects in the nanowires.

Using the tip as a movable gate allows us to individually control the charge on one dot in a nanowire that contains many dots, like the devices shown here. This movable gate technique has a great advantage over static gating techniques for two reasons. First, the movable gate allows one to image and locate the position of one or more quantum dots in a nanowire. Second, the tip can be used to address an individual dot in a nanowire with multiple dots; this can be difficult using lithographically defined gates if the dot locations are unknown, or if the spacing between two dots is smaller than the lithographic resolution.

The images in Figs. 3A-C show how the nanowire conductance is pinched off by negative voltages on the backgate to produce Coulomb conductance peak patterns characteristic of weakly coupled quantum dots [19]. A series of SPM images of device D1 are shown in Figs 3A-C for $V_{bg}$ = -1.94 V, $V_{bg}$ = -2.05 V and $V_{bg}$ = -2.12 V respectively. As the backgate voltage becomes more negative, the overall signal is reduced, and conductance only occurs when the tip is near the intersections of rings from different dots, where each dot is on a conductance peak. This is the expected pattern for multiple quantum dots in series with little coupling between them [19]. The small pink dots in Fig. 3 again denote dot locations and the dashed ellipses in Figs. 3B-C show the location of rings for the two outer dots taken from Fig. 3A. In addition to tuning the electron number on each dot, the backgate tunes the tunnel barriers forming the dots, making them more opaque and reducing the interdot coupling as $V_{bg}$ is made more negative. In the limit of completely decoupled dots in series, the condition for conductance is that all of the dots are on a Coulomb blockade peak. In conductance images like Figs. 3B-C, this condition occurs at the intersections of rings from different dots as spatially localized regions of peaked conductance. This is clearly seen in Fig. 3C.

The SPM images of device D2 in Figs. 4A-C show how the Coulomb blockade rings from a given dot evolve as tip voltage is increased from $V_{tip}$ = 0.48 V to 1.44 V. The rings move radially outward as $V_{tip}$ is increased, and their spacing decreases. In these images, one dominant set of rings is centered on the quantum dot in the upper half of the image. A black triangle is superimposed in the images to track the location of the Coulomb ring corresponding to the addition of the $n^{th}$ electron to the dot. The radius of this ring grows as $V_{tip}$ is increased, because more electrons are added to the dot by the positive tip voltage, as in Eq. 1. Thus the location of the $n^{th}$ Coulomb ring moves outwards from the quantum dot. In addition, the rings become more closely spaced as $V_{tip}$ is increased. This occurs because the tip-to-dot capacitance $C_{t-d}$ decreases with increasing tip to dot distance $r_{t-d}$ as expected for distances greater than the dot size.

In summary, a detailed picture of the potential landscape inside semiconducting nanowires has been revealed using our SPM imaging technique. This technique will be very useful for device development of future nano-electronic devices that employ semiconductor nanowires.

We acknowledge useful discussions with Jorden van Dam and Silvano de Franceschi. This work was supported at Harvard and at Delft by the Nanoscale Science and Engineering Center (NSEC), grant NSF PHY-01-17795, and at Delft by funding from NanoNed.

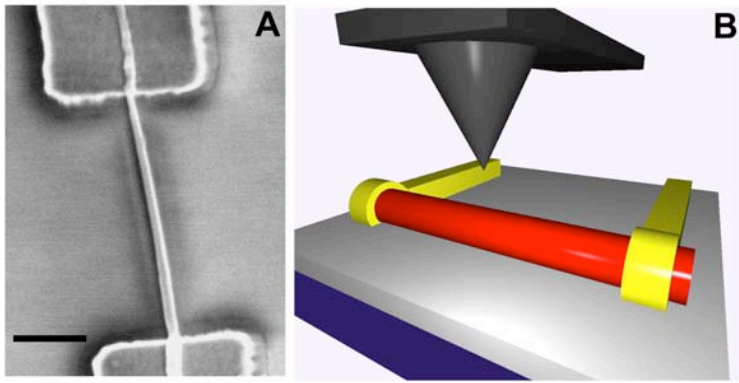

Figure 1. (A) SEM photo of an InAs nanowire (device D1) contacted with Ti/Al electrodes. (The slight kink in the wire at the top contact, due to AFM tip crash, occurred after the data presented in this paper was obtained). The scale bar is 500 nm long. (B) Imaging schematic. A charged AFM tip is scanned ~100 nm above the contacted InAs nanowire. Nanowire conductance as a function of lateral tip position is recorded to form an image. The wire lies atop a conducting Si substrate with a 250 nm thick $SiO_2$ capping layer.

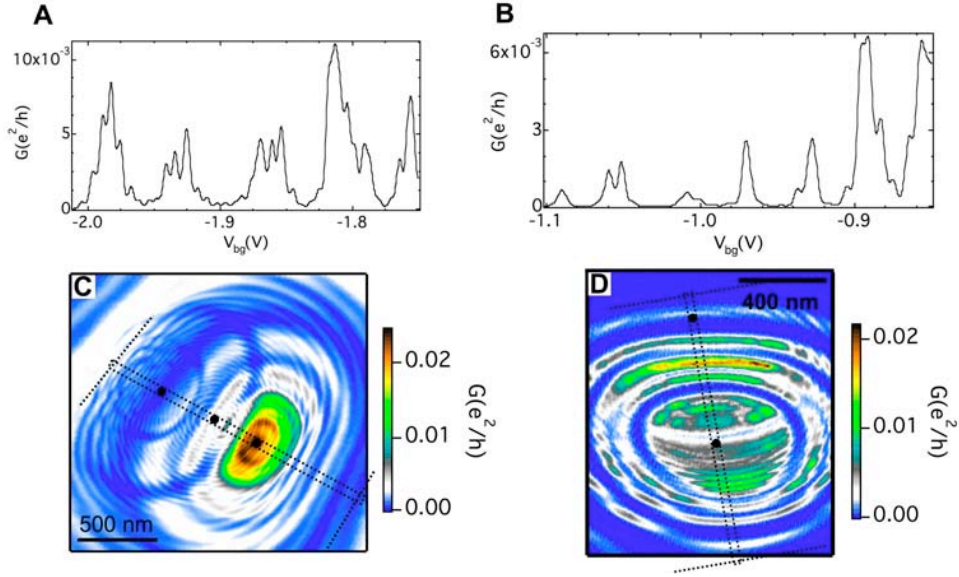

Figure 2. InAs nanowire transport measurements and corresponding images that spatially illuminate the behavior. (A-B) Nanowire conductance $G$ vs. backgate voltage $V_{bg}$ for devices D1 and D2 respectively. The plots show a complex pattern of Coulomb blockade conductance peaks characteristic of multiple quantum dots in series. From these plots, it is difficult to determine the number and locations of the dots in each wire. (C-D) SPM images of devices D1 and D2 respectively that display $G$ vs. position of a charged SPM tip scanned along a plane 100 nm above the nanowire. Concentric rings of high conductance, corresponding to Coulomb blockade peaks, are centered on quantum dots in the nanowire. (C) shows three sets of concentric rings identify three quantum dots whose positions are marked by black dots. (D) reveals rings surrounding two quantum dots in the nanowire. Dotted lines denote the outline of the wire and the electrical contacts.

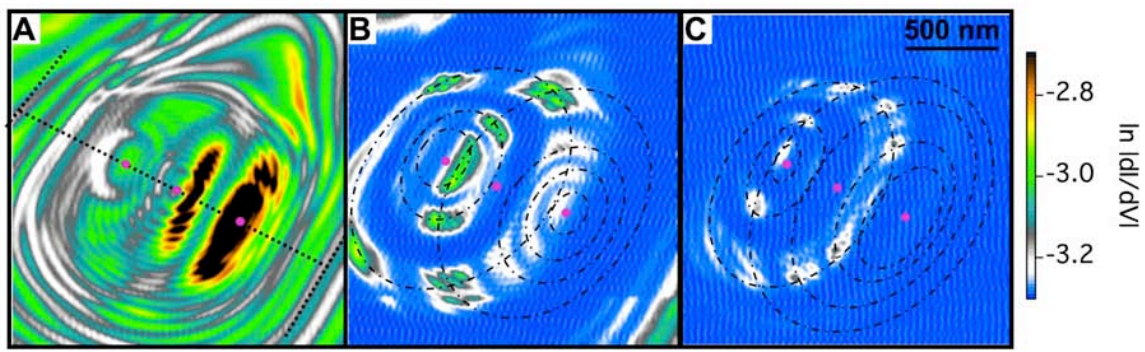

Figure 3. SPM images of conductance *G* for device D1 showing the interaction of Coulomb blockade rings from the three quantum dots in the nanowire. Pink dots mark the dot locations and dashed lines show outlines of the nanowire and contacts in (A). The images were recorded with $V_{tip} = 0$ V and backgate voltages (A) $V_{bg} = -1.94$ V, (B) -2.05 V and (C) -2.12 V. As $V_{bg}$ is made more negative, conductance occurs only near the intersection of rings from different dots, where each dot is on a Coulomb blockade conductance peak. Elliptical dash-dotted rings in (C) and (D) show the location of rings in (A) from the two outermost dots. The expected rings for the middle dot have not been shown, because they are so closely spaced that their inclusion would clutter the image.

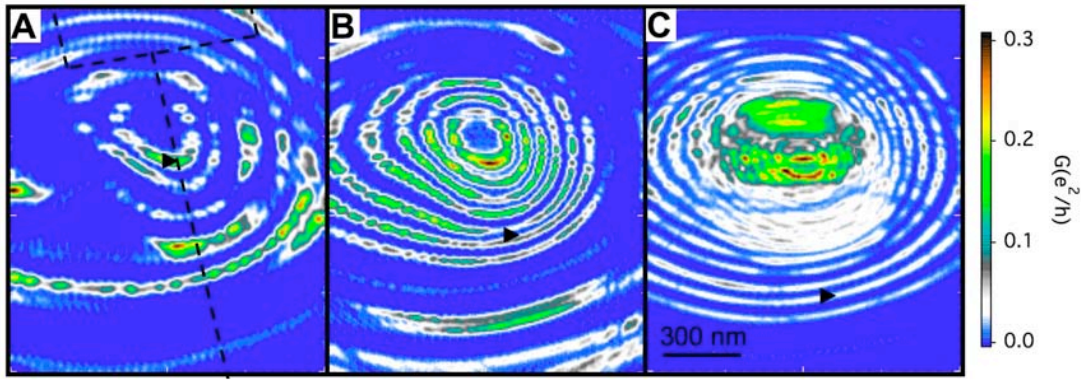

Figure 4. Evolution of SPM images of device D2 with tip voltages: (A) $V_{tip}$ = 0.48 V, (B) 0.90 V (C) 1.44 V. The wire and top contact are denoted with dashed lines in (A). Coulomb blockade rings surround a quantum dot in the upper half of the image. As $V_{tip}$ increases, the rings expand outwards in size and become more closely spaced. The black triangle tracks one Coulomb peak, demonstrating how the size of the rings grows with tip voltage.

REFERENCES

(1) Lieber, C. M. *Mat. Res. Soc. Bull.* **2003**, *28*, 486.

(2) Yang, P. *Mat. Res. Soc. Bull.* **2005**, *30*, 85.

(3) Samuelson, L.; Thelander, C.; Bjork, M.T.; Borgstrom, M.; Deppert, K.; Dick, K.A.; Hansen, A.E.; Martensson, T.; Panev, N.; Persson, A.I.; Seifert, W.; Skold, N.; Larsson, M.W.; Wallenberg, L.R. *Physica E*: *Low-dimensional Systems and Nanostructures* **2005**, *25*, 313.

(4) Zhong, Z.; Fang, Y; Lu, W.; Lieber, C.M *NanoLetters* **2005**, *5*, 1143.

(5) De Franceschi, S.; van Dam, J. A.; Bakkers, E. P. A. M.; Feiner, L. F.; Gurevich, L.; Kouwenhoven, L. P. *Appl. Phys. Lett.* **2003**, *83*, 344.

(6) Bjork, M. T.; Thelander, C.; Hansen, A. E.; Jensen, L. E.; Larsson, M. W.; Wallenberg, L. R.; Samuelson, L. *NanoLetters* **2004**, *4*, 1621.

(7) Xiang, J.; Lu, W.; Hu, Y.; Wu, Y.; Yan, H.; Lieber, C.M. *Nature* **2006**, *441*, 489.

(8) Doh, Y.; Van dam, J. A.; Roest, A. L.; Bakkers, E. P. A. M.; Kouwenhoven, L P.; De Franceschi, S. *Science* **2005**, *309*, 272.

(9) Topinka, M. A.; Westervelt, R. M.; Heller, E. J. *Phys. Today* **2003**, December, 47.

(10) Woodside, M. T.; McEuen, P. L. *Science* **2002**, 296, 1098.

(11) Fallahi, P.; Bleszynski, A. C.; Westervelt, R. M.; Huang, J.; Walls, J. D.; Heller, E. J.; Hanson, M.; Gossard, A. C. *NanoLetters* **2005**, 5, 223.

(12) Pioda, A.; Kicin, S.; Ihn, T.; Sigrist, M.; Fuhrer, A.; Ensslin, K.; Weichselbaum, A.; Ulloa, S. E.; M. Reinwald, M.; Wegscheider, W.; *Phy. Rev. Letters* **2004**, *93*, 216801.

(13) Bockrath, M.; Liang, W.; Bozovic, D.; Hafner, J. H.; Lieber, C. M.; Tinkham, M.; Park, H. *Science* **2001**, *291*, 283.

(14) Tans, S.J.; Dekker, C. *Nature* **2000**, *404*, 834.

(15) Zhitenev, N. B.; Fulton, T. A.; Yacoby, A.; Hess, H. F.; Pfeiffer, L. N.; West, K. W. *Nature* **2000**, *404*, 473.

(16) Ahn, Y.; Dunning, J.; Park, J. *Nano Letters* **2005**, *5*, 1367.

(17) Gu, Y.; Kwak, E.-S.; Lensch, J. L.; Allen, J. E.; Odom, T. W.; Lauhon, L. J. *Applied Physics Letters* **2005**, *87*, 043111.

(18) Bakkers, E. P. A. M.; Van Dam, J. A.; De Francheschi, S.; Kouwenhoven, L.P.; Kaiser, M.; Verheijen, M.; Wondergem, H.; Van Der Sluis, P. *Nature Materials* **2004**, *3***,** 769.

(19) Waugh, F. R.; Berry, M. J.; Mar, D. J.; Westervelt, R. M.; Campman, K. L.; Gossard, A. C. *Phys. Rev. Lett*. **1995**, *75*, 705.

(20) Kouwenhoven, L. P.; Marcus, C. M.; McEuen, P. L.; Tarucha, S.; Westervelt, R. M.; Wingreen, N. S.; in *Mesoscopic Electron Transport;* Sohn, L. L.; Kouwenhoven, L. P.; Schon, G., Eds.; Kluwer: Dordrecht, 1997.

(21) Ruzin, I. M.; Chandresekar, V.; Levin, E. I.; Glazman, L. I. *Phys. Rev. B* **1992**, 45, 13469.